# Searching Data: A Review of Observational Data Retrieval Practices in Selected Disciplines

Kathleen Gregory[1], Paul Groth[2], Helena Cousijn[3], Andrea Scharnhorst[1], Sally Wyatt[4]

[1]Data Archiving and Networked Services, Royal Netherlands Academy of Arts and Sciences;
[2]Elsevier Labs; [3]Elsevier, Datacite; [4]Faculty of Arts and Social Sciences, Maastricht University

## Abstract

A cross-disciplinary examination of the user behaviors involved in seeking and evaluating data is surprisingly absent from the research data discussion. This review explores the data retrieval literature to identify commonalities in how users search for and evaluate observational research data in selected disciplines. Two analytical frameworks, rooted in information retrieval and science and technology studies, are used to identify key similarities in practices as a first step toward developing a model describing data retrieval.

## Introduction

Open research data is touted as having the potential to transform science and fast-track the development of new knowledge (Gray, 2009). In order for data to fulfil this potential, users must first be able to find the data that they need. This is not a simple task. Facilitating data discovery relies on developing underlying infrastructures, support systems, and data supplies (Borgman, 2015). It is equally important to understand the behaviors involved in data retrieval, but a user-focused, cross-disciplinary analysis of data retrieval practices is lacking. This review explores the existing data retrieval literature and identifies commonalities in documented practices among users of observational data as a first step towards creating a model describing how users search for and evaluate research data.

Although information retrieval (IR) has been extensively studied for over sixty years (Sanderson & Croft, 2012), data retrieval is a nascent field. Recent studies surrounding the issue examine how data are made available via data sharing (Tenopir et al., 2011, 2015), how researchers reuse data (Faniel, Kriesberg, & Yakel, 2016; Pasquetto, Randles, & Borgman, 2017), and how systems are designed to optimize data discoverability and retrieval (Pallickara, Pallickara, & Zupanski, 2012). Information documenting data retrieval behaviors is buried throughout other disciplinary and data-related literature and is not easy to identify (Gregory et al, 2018).

We draw on work in information retrieval and science and technology studies (STS) to guide the identification of this buried literature and to develop our analysis frameworks. The first framework is based on established models of interactive information retrieval; the second framework builds on STS-inflected work examining data practices and communities. We begin by discussing the frameworks in more detail before using them to present and synthesize the data retrieval behaviors documented in the collected literature. We end with a discussion of commonalities across disciplinary communities and identify gaps in the literature and areas for future work.

## Framework #1: A Broad View of Interactive Information Retrieval

Information retrieval is an interactive process, involving a dynamic interplay between users and IR systems (Xie, 2008). Numerous models describe user-oriented interactive IR. Three of the most pivotal are Ingwersen's cognitive model (Ingwersen, 1992, 1996), Belkin's episode model (Belkin, 1993, 1996), and Saracevic's stratified interaction model (Saracevic, 1996, 1997). Detailed characterizations of the strategies (e.g. Bates, 1990) and cognitive and affective stages in user-oriented information seeking (Kuhlthau, 1991) have also been proposed. Despite their differences, established models assume that users are actively involved in the search process and that context influences search behaviors (Rieh & Xie, 2006; Xie, 2008).

Interactive IR models share a few key stages[1] (Wolfram, 2015) that are used to structure the first framework and to provide the main divisions of this paper:

- **Users and Needs**: describes user contexts and data needs
- **User Actions:** describes the sources and search strategies used to locate research data
- **Evaluation:** describes criteria and processes used when evaluating data for reuse

The term data retrieval is used in this review to refer to this entire complement of needs, actions, and evaluation behaviors.

## Framework #2: A Broad View of Data Communities

Data practices can define communities in different ways (Birnholtz & Bietz, 2003). Data communities form around disciplinary domains, (Faniel, Kansa, Kansa, Barrera-Gomez, & Yakel, 2013; Palmer, Cragin, & Hogan, 2004), research approaches and data collection methodologies (Birnholtz & Bietz, 2003; Weller & Monroe-Gulick, 2014), and particular data sources (Brown, 2003; Sands, Borgman, Wynholds, & Traweek, 2012). Both macro-level characteristics, such as using quantitative vs qualitative data (Birnholtz & Bietz, 2003) and micro-level characteristics, such as participation in a specific research project, (Borgman, Wallis, & Enyedy, 2007) can define community membership. A researcher may belong to multiple data communities simultaneously, or she may choose to define her community in unique ways (Birnholtz & Bietz, 2003).

Here, we embrace a broad approach to conceptualizing data communities. The overarching data community used in this framework is based on accepted classifications of research data. While classifying data is a notoriously difficult task (Borgman, 2015), broad categories that have proven to be useful are observational, experimental, or computational data (National Science Board, 2005; National Science Foundation, 2007). As a first step in testing the validity of this conception of data communities, we focus on a community bounded by the use of a particular data type: observational data.

Observational data result from recognizing, recording, or noting occurrences. They are often produced with the help of instruments, and include weather observations, polling data, photographs, maps, and economic indicators (Borgman, 2015; National Science Board, 2005). Observational data are used across disciplines; we therefore introduce disciplinary communities into the second framework to provide another level of analysis.

Science and technology studies research explores the role of disciplinary norms and behaviors in data practices (e.g. Leonelli, 2016). Subdisciplines and individual research groups may have unique data practices, different than those of the broader disciplinary community (Gregory et al, 2018); while these differences are important, we suggest that commonalities are also important. In order to identify possible commonalities, we group the disciplines represented in the retrieved literature into five broad domains: astronomy, earth and environmental sciences (EES), biomedicine, field archaeology and social sciences.

This review centers on the role of the researcher as data user. While the discussion of data communities often takes the perspective of data producers, researchers play multiple roles, often mixing data production and consumption (Borgman, Van de Sompel, Scharnhorst, van den Berg, &

Treloar, 2015). We focus on consumers/users of observational data who use data they did not create either for new purposes and/or to support existing projects.

## Purpose of the Frameworks

Many studies employ case studies, interviews and ethnographic research to depict particular data practices in fine detail (Cragin, Chao, & Palmer, 2011; Weber, Baker, Thomer, Chao, & Palmer, 2012) and are spread across disciplinary domains. While these studies provide great depth, it is challenging to bring them together in meaningful ways to identify similarities (Faniel, Barrera-Gomez, Kriesberg, & Yakel, 2013). The primary goal of this review is to use the macroscopic perspectives of the frameworks introduced above to identify commonalities in reported practices. Such a broad approach comes with two drawbacks: the loss of some of the complexity and detail of the original studies and a bias in the disciplinary scope.

Each section begins with a table synthesizing the reviewed literature through the lens of both frameworks. We then present the literature used to create these syntheses, structuring the findings by disciplinary community. In the discussion, we summarize and discuss the key findings from each section and identify common themes.

## Methodology

Our literature collection methodology was informed by the first framework. We performed keyword searches related to information retrieval (e.g. user behavior, information seeking) and data practices (e.g. data sharing, data reuse, research practices) across all fields, primarily in the Scopus database. We also performed searches related to data search and data discovery and used bibliometric techniques such as citation chaining and related records.[2]

We closely read the nearly 400 retrieved documents to identify papers referring to observational data. As we read, we again applied the first framework, seeking descriptions of data users and their needs, sources and strategies used to locate data, and the criteria used to evaluate data for potential reuse. Few studies examine data retrieval practices directly; much of the information is buried within investigations of data sharing and data reuse or found in user studies of particular repositories.

## Users and Needs

In this section we analyze the diversity of users' data needs within the context of disciplinary communities. We adopt the characterization of background uses of data which support research and foreground uses which drive new research (Wynholds et al., 2012).

| | Needs | |
|---|---|---|
| Users in this community… | need this type of data | for these purposes (*italicized=foreground*, normal=background) |
| Astronomy | Data from sky surveys, telescopes, archives, repositories, data catalogs, virtual observatory systems | *New questions of old data*, baselines, instrument calibration, physical properties, model inputs, data integration |
| Earth & Environmental Sciences | Plant, animal, water, weather, solar observations; soil analyses, rock thin-section and satellite images; maps, geographic, demographic and census data; continuously collected and transmitted data, data at temporal/spatial scales, raw and summarized data | *New questions of old data, meta-analyses*, calibration, context, baselines, reference, model inputs, verification, comparison, environmental planning, policy- and decision making, education, instrument monitoring; data integration |

| Biomedicine | Images, complete fMRI studies, pathology results, patient observations and demographics; population-level disease data, behavioral data | *Disease/disorder research, new visualizations*, evaluations, 3-D anatomical pictures, preparing research outputs, education, patient care |
|---|---|---|
| Field Archaeology | Field notebooks, photographs, artefacts, stratigraphic baselines; data at temporal/spatial scales | *New insights from data aggregation*, comparison, triangulation; training, dissertations, assignments, preparing tours, inventories of local excavations |
| Social Sciences | Survey data (often only one question is of interest), long-running datasets/surveys, interviews, archival documents, images, videos | *Re-interpret datasets; new questions, comparative research*, comparison, preparations, training, dissertations |

*Table 1*: Users' observational data needs by disciplinary community.

**Astronomy**
Much astronomical research can be classified as big science, involving large international projects supported by extensive knowledge sharing infrastructures (Borgman et al., 2007). Big science is not the only approach, as astronomers also conduct research falling within the long tail of science (Wynholds et al., 2011). Access to the vast amount of available research data is remarkably open, and data sharing is generally encouraged (Hoeppe, 2014; Pepe, Goodman, Muench, Crosas, & Erdmann, 2014).

**Data needed.** Data from large-scale sky surveys, such as the Sloan Digital Sky Survey (SDSS), form the foundation for many research projects (Pepe et al., 2014). Similarly, the data practices of researchers working with the SDSS are the cornerstone of the data retrieval literature in astronomy (Borgman et al., 2016; Borgman, Darch, Sands, & Golshan, 2016; Sands et al., 2012; Wynholds et al., 2011).

Sky survey data fuel studies involving further data processing; derived data are then used as the basis for publications (Pepe et al., 2014). Direct data from ground- and space-based telescopes, data located in data repositories and catalogs, and data found through federated queries of virtual observatory systems are important sources (Sands et al., 2012; Wynholds, Wallis, Borgman, Sands, & Traweek, 2012). Theoretical researchers also use observational data from established archives as model inputs (Sands et al., 2012).

**Data uses**. Astronomers combine multiple datasets, often from multiple archives or telescope types, during a single project (Sands et al., 2012; Wynholds et al., 2011). Merging data about the same target from different instruments poses a significant challenge (Hoeppe, 2014; Zinzi at al., 2016).

Astronomers use external data for foreground purposes driving new scientific inquiries and leading to new discoveries (Wynholds et al., 2012; Wynholds et al., 2011), and for background purposes supporting research, such as study baselines, calibrating instruments, and searching for specific physical properties (Wynholds et al., 2012).

**Earth and Environmental Sciences**
A variety of disciplines and subdisciplines are represented in the literature at differing levels of granularity. Data retrieval practices are sparsely documented in fields such as volcanology, but discussions are increasing in other disciplines, i.e. the water sciences (e.g. Dow, Dow, Fitzsimmons, & Materise, 2015). This is partly due to a change in data collection techniques. As researchers transition from primarily manual field work to using sensors enabling continuous collection, they must find new ways to manage their data (Maier et al., 2012). The ecologists involved in the multidisciplinary Center for Embedded Networked Sensing (CENS) are an example of researchers caught in this transition (see Borgman et al., 2007; Wallis, Rolando, & Borgman, 2013).

**Data needed.** Biodiversity researchers require an incredible multiplicity of data. Potentially any information about life on earth, from satellite photos to forest inventories, could be important (Bowker, 2000b). Scientists need information about species distribution and occurrence, population trends, and geographic raw data (Davis, Tenopir, Allar, & Frame, 2014). The needs of CENS researchers exemplify what Bowker terms "data diversity," as they use weather, solar, and river observations, as well as remote sensing and demographic data (Bowker, 2000a; Wallis, et al., 2013). Data diversity is also the norm in the geo- and water sciences. Volcanologists rely on images of thin rock sections, chemical analyses and characterizations of the earth's crust. Additionally, stratigraphers use astronomical observations and numerical data extracted from graphs to study geologic history (Weber et al., 2012). Geographers need data spanning the physical and social sciences, requiring topographic, geologic and demographic maps, satellite images and drawings, and census data (Borgman et al., 2005). Water scientists need streamflow, evaporation, groundwater level, and water quality measurements (Beran, Cox, Valentine, Zaslavsky, & McGee, 2009). Although they do not exist for every condition, continuously collected data that can be analyzed by location and time are expected (Dow et al., 2015).

This need for data at different geographic and temporal scales connects the disciplines. Atmospheric scientists need large amounts of observational data from specific regions and times for their models (Pallickara et al., 2012). Data collected at local levels can be more important than data collected at national or state levels, as shown by a user survey from (Davis et al., 2014).

The Davis et al. survey is one of the few that differentiates between the data needs of different types of users; another example is a study at the Center for Coastal Margin Observation and Prediction (CMOP) (Maier et al., 2012). Internal and external researchers using CMOP data want succinct data overviews. Policy and decision makers need thematic collections summarized on one page, with salient data clearly marked; users in education sectors are also interested in CMOP data, although their specific needs have not yet been studied (Maier et al., 2012).

Like researchers, environmental policy and decision makers need information from different locations and times, but they have difficulties accessing the information (McNie, 2007) or finding the right type. Data produced by scientists are not automatically useful for policy makers (Cash et al., 2003). Environmental planners, i.e., may not need the same depth of information as researchers (Van House, Butler, & Schiff, 1998); reflecting this, differentiated data products for diverse users are being explored (see Baker, Duerr, & Parsons, 2015).

**Data uses.** CENS researchers use external data solely for background purposes, such as contextualizing their own data and calibrating instruments (Wallis et al., 2013; Wynholds et al., 2012). Other background uses include benchmarking and as references (Bowker, 2000b). Some ecologists do reuse external data to answer new questions (Zimmerman, 2007) or to create meta-analyses (Michener, 2015).

Integrating diverse data is problematic across the environmental sciences. Data collected at different scales and using different nomenclatures are difficult to merge (Dow et al., 2015; Maier, et al., 2014; Bowker, 2000b). Natural variances in systems and populations further complicate fitting biodiversity data together (Bowker, 2000b; Zimmerman, 2007). Stratigraphers use one dataset to calibrate another as they construct geologic timelines used as baseline data by other researchers (Weber et al., 2012). Atmospheric scientists and climatologists grapple with problems stemming from metadata variation (Pallickara et al., 2012) and differences in community data practices (Edwards, Mayernik, Batcheller, Bowker, & Borgman, 2011).

Modelers use external data at specific points in the research process. After reformatting and regridding data to fit model specifications, earth scientists use observational data to initially force models and for parameterization; data availability limits the types of studies undertaken (Parsons, 2011). Coastal modelers engage in similar behavior, continually calibrating and benchmarking their models, and comparing outputs to external observational data (Maier et al., 2012; Weber et al., 2012).

Environmental planners use data not only to make decisions, but also to defend their viewpoints, to persuade, and in education. (Van House et al., 1998). Although detailed studies of non-scientists' data needs are lacking (Faniel & Zimmerman, 2011), reported "background uses" of oceanographic data include preparation for triathlons, search and rescue operations or fishing expeditions (Weber et al., 2012)

**Biomedicine**

The biomedical literature focuses on fields centering on imaging, such as neuroscience and radiology.

**Data needed.** As neuroscience embraces big science methodologies, the field is struggling with how to make data available, discoverable, and usable (Choudhury, Fishman, McGowan, & Juengst, 2014). Researchers rely on visualizations of normal and abnormal brains, although they also consult brain bank samples (Beaulieu, 2004). Sometimes researchers need raw fMRI studies, including detailed metadata; sometimes images and scans suffice (Key Perspectives, 2010; Van Horn & Gazzaniga, 2013). Neuroimaging data are complex, consisting of numerous brain section slices, time-points, and other variables (Honor, Haselgrove, Frazier, & Kennedy, 2016). Neuroscientists incorporate more than just imaging into their work, using demographic, genetic, and behavioral data (Williams et al., 2009).

Clinicians and medical researchers also use a mixture of images and other observational data, such as pathology results, clinical data (e.g. progression of tumor grades), patient demographics, and population-level disease data (Kim & Gilbertson, 2007). Medical images are an essential part of workflows in fields such as radiology (Markonis et al., 2012), where health care professionals tend to search for two types of images: general medical images (e.g. images of anatomic organs) and specific medical images, which are used for clinical or comparison purposes (Sedghi, Sanderson, & Clough, 2011). Users need images collected with different modalities (X-rays, CT scans, and MRIs) (Kim & Gilbertson, 2007); medical students need images corresponding to their current courses (Müller et al., 2006). All reusable medical data must be provided in a way protecting patient privacy (Erinjeri, Picus, Prior, Rubin, & Koppel, 2009).

**Data uses.** Neuroscientists use imaging data for comparisons, evaluations, and creating three-dimensional pictures of brain anatomy (Beaulieu, 2004). A single scan is of little value unless incorporated into a larger database of scans. Aggregating individual scans creates complete virtual brains that can be manipulated to facilitate new discoveries (Beaulieu, 2004), as in the case of combining fMRI scans from different populations to yield insights about Alzheimer’s biomarkers. (Van Horn & Gazzaniga, 2013).

In a study of clinicians, researchers, educators, librarians, and students, users incorporate images in research, patient care and education (Hersh, Müller, Gorman, & Jensen, 2005). A follow-up study further characterizes these needs, showing that images are used for self-education; educating medical students, patient education, making difficult diagnoses, and developing research ideas, grant proposals, and publications (Kalpathy-Cramer et al., 2015).

**Field Archaeology**

Archaeology is another field in transition. Methodologies and data practices are changing, as data move away from being published in analog-only formats to being made available in digital repositories (e.g. Arbuckle et al., 2014); this facilitates data aggregation to study phenomena such as domestic livestock expansion (Arbuckle et al., 2014; Atici et al., 2017). Interdisciplinarity and data diversity are thriving in archaeology, as research projects can involve soil scientists, zooarchaeologists, and material scientists (Faniel, Kansa, et al., 2013)

Metadata and documentation of methods and site conditions are extremely important in archaeology, as original sites are often “decomposed” during the research process (Faniel, Kansa, et al., 2013). Data recording and metadata standards do not exist (Faniel & Yakel, 2017); making integration across contexts and collection methodologies challenging (Niccolucci & Richards, 2013; Faniel & Yakel, 2017).

Field archaeologists need field notes, photographs, and artefacts in museum collections (Faniel, Kansa, et al., 2013). Geographic, stratigraphic and chronological baseline data are also vital (Atici, Kansa, Lev-Tov, & Kansa, 2013). Archaeologists compare finds from the field to museum collections, often triangulating data from multiple sources (Faniel, Kansa, et al., 2013). Researchers are not the only “consumers” of archaeological data; students, hobbyists, and employees of museums and companies use data for diverse background and fewer foreground purposes, e.g. aggregating discrete units of “raw data” (Borgman, Scharnhorst, & Golshan, 2018).

**Social Sciences**

Reusing quantitative data in the social sciences is well-established (Kriesberg et al., 2013; Faniel & Yakel, 2017); the reuse of qualitative data is complicated by issues of participant confidentiality and the embeddedness of the researcher in data creation (Broom, Cheshire, & Emmison, 2009).

Social scientists need data from surveys and long-running datasets (Shen, 2007). Researchers are often interested in only one data point or survey question. Details about the operationalized variables or measured constructs usually are not present when examining individual questions in isolation (Dulisch, Kempf, & Schaer, 2015). Social scientists also need archival documents, images, videos, and interview data (Karcher, Kirilova, & Weber, 2016).

Data can be reused for comparative research or to ask new questions, reinterpret datasets, or verify findings (Corti, 2007). Background uses, i.e. preparing for data collection, are common (Parry & Mauthner, 2005).

Kriesberg and colleagues examine the needs of early career researchers (ECRs) in quantitative social sciences, archaeology and zoology. External data are used in training and dissertations; young researchers may reuse data more often, due to difficulties collecting their own data (2013).

## User Actions

This section examines the resources and strategies used within different communities to locate data.

| | Actions | |
|---|---|---|
| Users in this community… | use these resources | in this way |
| Astronomy | NASA archives, journals, personal exchanges, personal websites, general search engines | Querying archives, extracting data from articles into new tables, informal personal requests |
| Earth & Environmental Sciences | Journals, personal exchanges, repositories, databases, natural history collections, general search engines, industry | Extracting data from articles, email/ telephone/letters, metadata searches, faceted searching, filtering, aggregating data to create new datasets, "bounded" strategies (by journal, location, time) |
| Biomedicine | Online image repositories, local image and patient information systems, personal image collections, Google Images, journals | local systems - patient name/identifier; Online sources - keyword and hierarchical searches, short queries for images |
| Field Archaeology | Personal connections – museum staff and data producers, natural history collections, museums, repositories/archives, publications | Searching by location (keywords, browsing), collaborations to gain additional data |
| Social Sciences | Survey banks, data catalogs (i.e. DBK), repositories, governmental/ statistical offices, databases, commercial providers, personal connections, publications | Following publication references; survey banks - short queries, mismatch between strategies and database design, DBK - more time spent than in literature searching, keyword searching followed by browsing, filters and author names not used, |

*Table 2*: Actions taken to locate data

**Astronomy**
Astronomers are generally efficient information seekers, in part due to strong disciplinary infrastructures and tools (Meyer et al., 2011). SDSS users download data directly from NASA archives or obtain them from public data releases (Sands et al., 2012). Discovering and tracking down smaller datasets is challenging; SDSS users sometimes browse personal websites or use general search engines. They then contact research groups directly with their data requests. Despite well-developed infrastructures, personal networks remain an important means for identifying and obtaining data (Sands et al., 2012).

Journal articles are another important data source. Astronomers copy and paste or transcribe data from articles into new tables for further manipulation (Pepe et al., 2014). Direct citation of archival accession numbers facilitates data discovery from journals (Swan & Brown, 2008).

**Earth and Environmental Sciences**

Finding and accessing biodiversity data can be challenging, although academics have an easier time than government employees and program managers. A lack of training, time, and knowing where to look hinders effective data search among these groups (Davis et al., 2014). Knowing where to search can be especially problematic in areas outside of a researcher's primary expertise (Devarakonda, Palanisamy, Green, & Wilson, 2011) and is contingent on knowing that data even exist (Zimmerman, 2003). Personal experiences with data collection and a familiarity with research trends help researchers estimate whether data are extant and findable (Zimmerman, 2007).

Compounding this problem, data are distributed across numerous repositories (Dow et al., 2015). Users must first discover the repository, and then invest significant time and energy becoming familiar with each search environment (Ames et al., 2012; Beran et al., 2009). Given the diversity of interfaces, it is not surprising that water scientists desire a "Google for data" (Megler & Maier, 2012).

In a global survey of the environmental research community, the majority of respondents discover data through journal articles, search engines, and disciplinary repositories; 40% request data directly from data providers (Schmidt, Gemeinholzer, & Treloar, 2016). Although some environmental planners are interested in using journals and primary sources, they find it too time-consuming (Miller et al., 2009), and may instead turn to colleagues for biodiversity information (Janse, 2006; Pullin, Knight, Stone, & Charman, 2004).

Stratigraphers extract data from journals, laboriously re-creating tables from published graphs. They are willing to spend money as well as time obtaining data, sometimes purchasing expensive high-resolution data from drilling companies (Weber et al., 2012). Geographers utilize journals and search engines to locate maps, images, and repositories, but poor indexing and metadata derail their efforts (Borgman et al., 2005). Ecologists in Zimmerman's studies gather single data points from multiple sources and then aggregate them to create new datasets (2007; 2008), an approach that is increasingly common in biodiversity research (Davis et al., 2014).

Personal exchanges are valuable, if complex, sources of external data. Requesting data from CENS, for example, is a multistep process. Data seekers identify CENS as a potential source, contact the CENS researcher, and discuss the availability and suitability of the data. The CENS researcher then gathers, processes, and delivers the requested data (Wallis et al., 2013). Ecologists employ a variety of tactics (emails, letters, and telephone calls) to obtain data mentioned in articles. As organizations grow and such requests increase, personal exchanges cease to be an effective way to obtain data (Wallis et al., 2007).

Ecologists reusing data employ "bounding" strategies, limiting searches to particular journals, times or locations to collect representative samples (Zimmerman, 2007). As data seeking is data collection, these researchers use strategies that minimize error, can be publicly defended, and increase the likelihood of accessing data (Zimmerman, 2007). They have specific search criteria; the general information in databases usually does not meet their detailed needs (Zimmerman, 2007). Before building specific search tools, CMOP researchers struggled with similar problems, retrieving either zero or thousands of hits. If researchers found searching too frustrating, they would simply stop searching (Maier et al., 2012; Megler & Maier, 2012).

Large atmospheric datasets, encoded in binary formats to facilitate storage and transfer, cannot effectively be searched with text-based search engines. Rather, users must browse collections using metadata schemas (Pallickara, Pallickara, Zupanski, & Sullivan, 2010). For other data, i.e. datasets in the DataONE platform, users prefer keyword searches, followed by filtering (Murillo, 2014).

**Biomedicine**

While it has become easier to locate data, for example in neuroscience (Beaulieu, 2004), access restrictions still frustrate researchers (Honor et al., 2016)

Medical image retrieval studies show that users search both local restricted-access systems and free internet sources. Local systems, including Picture Archiving and Communication Systems (PACS), electronic patient records, hospital archives, and teaching files, house images and patient data (Müller et al., 2006). Radiologists also curate their own collections of images stored on personal computers (Markonis et al., 2012).

Despite access to specialized collections, internet searches, particularly with Google Images, are common (Markonis et al., 2012; Müller et al., 2006). Limitations of such searches include sifting through irrelevant results and a dearth of highly-specialized images. Nevertheless, online image repositories are unpopular among health care professionals, perhaps because of their limited scope (Sedghi et al., 2011). Academic journals, however, facilitate locating specialized, cutting-edge images with contextual information that are difficult to locate on the web (Sedghi et al., 2011).

Search strategies vary depending on the searcher's professional role, although commonalities do exist. Users often search by patient names or identifier in PACS for diagnostic purposes; brief keyword or hierarchical searching is typical in non-diagnostic searching (De-Arteaga et al., 2015; Markonis et al., 2012; Müller et al., 2006).

Success is not assured when searching for images. In a study of radiologists, users fail to find desired images in almost 25% of cases. Users believe these images exist, but that they simply cannot be found (Markonis et al., 2012). Possible search difficulties stem from a lack of time and available relevant papers, the newness of certain topics, and a lack of domain-specific search tools (Sedghi et al., 2011).

**Field Archaeology**

Data discovery is a significant problem in field archaeology. Data are scattered among collections or sometimes are only in unpublished field reports (Niccolucci & Richards, 2013). Although publications are used in data discovery (Faniel & Yakel, 2017), they do not consistently include data; a significant delay between data collection and publication exacerbates the problem (Kriesberg et al., 2013). Researchers often do not know what data are available (Aloia et al., 2017). ECRs circumnavigate difficulties by collaborating with supervisors to locate data (Kriesberg et al., 2013). Other archaeologists turn to personal networks, museums, and, as the shift toward digital data continues, data archives (Faniel, Kansa, et al., 2013; Faniel & Yakel, 2017). Details about how users search archives are sparse (Borgman et al., 2015), although searching and browsing by location are important strategies often complicated by differences in geographic terminology (Borgman et al., 2018).

**Social Sciences**

Social scientists use data from governmental/statistical offices and specialized databases (Shen, 2007). Economists also obtain data from statistical offices but may purchase data directly from commercial providers (Bahls & Tochtermann, 2013). Researchers easily locate data from national, publicly funded datasets, but struggle to locate smaller datasets and video data for reuse (Key Perspectives, 2010). Researchers tap publications or make direct requests to find this more specialized data (Swan & Brown, 2008).

Personal networks, including advisors, co-workers of advisors or former employers, are key sources of qualitative data (Yoon, 2014b), especially for ECRs, who rely on journal recommendations from advisors and observations of their colleagues (Kriesberg et al., 2013; Faniel & Yakel, 2017). Not knowing who to contact or where to begin searching makes locating relevant data difficult (Curty, 2016).

Searchers of the DBK, the primary catalog for social science data in Germany, expend more time and effort when seeking datasets than they do for publications. These researchers do not frequently use author names; rather, keyword searching, followed by browsing long results lists, are more frequent strategies. Researchers complain about a lack of filtering options, but do not use available filters (Kern & Mathiak, 2015). Social scientists search a survey bank by short keyword queries or social construct, even though these strategies do not match the database's structure (Dulisch et al., 2015).

## Evaluation

We identify major frames used in the literature to discuss data evaluation criteria, including trust, quality, necessary contextual information, and relevance. The frames overlap, as the characteristics composing these frames vary from paper to paper, both within and across disciplines. In the table below, we present the evaluation criteria and associated frames as they are discussed in the literature.

| | Evaluation |
|---|---|
| Users in this community… | use these criteria to evaluate data |
| Astronomy | **1. Contextual Information**: instrumentation, observational conditions, data processing, original research questions<br>**2. Trust:** author reputation, source reputation |
| Earth & Environmental Sciences | **1. Contextual Information:** instrumentation, observational conditions, data collection procedures, data processing, provenance, original research questions<br>**2. Quality:** meet community standards, comprehensiveness/continuity over time, estimations and uncertainties, resolution<br>**3. Trust:** source, knowledge of object and data collector, author reputation/affiliation, funder, community membership<br>**4. Understandability:** familiarity with practices, data type, subject; consult experienced researchers, first decode data<br>**5. Ease of access** |
| Biomedicine | **1. Quality**: noise, resolution, anatomical coverage, image acquisition details<br>**2. Trust:** supporting documentation, social networks<br>**3. Relevance**: experience, combination of textual/visual/medical criteria, visual relevancy, background information, understandability, image quality, modality, source |
| Field Archaeology | **1. Contextual information**: collection methods, instrumentation, observational conditions, provenance, original research goals, baseline geographic/stratigraphic/chronological data<br>**2. Suitability for analysis:** consistent data recording practices<br>**3. Trust**: reputation/affiliation/skill of authors, repository features, language in supporting documentation |
| Social Sciences | **1. Contextual Information**: collection methods, instrumentation, other analyses, definition/measurement of variables, data handling/processing<br>**2. Quality**: completeness, accessibility, ease of use, credibility, reputation of repository, reputation of author/journal **not** important<br>**3. Relevance:** time frame of study, keywords, citing literature, title and publication year **not** as important<br>4. **Trust**: prior reuse, reputation of data repository, reputation of data producer |

*Table 3*: Evaluation criteria with frames used in the literature

**Astronomy**
Astronomers rely on detailed documentation of instrumentation, collection methods and conditions, data processing, and original research questions (Borgman et al., 2016; Wynholds et al., 2011). They know which authors to trust and believe data in NASA archives and established projects are valid, accurate, and trustworthy. Researchers must completely understand data and the creation processes; they would rather recreate data before using poorly documented secondary data products (Wynholds et al., 2011).

**Earth and Environmental Sciences**
When evaluating data for reuse, researchers use contextual information about data provenance (Dow et al., 2015; Murillo, 2014), technical instrumentation (Wallis et al., 2007), and original research questions (Zimmerman, 2008). Researchers reuse data they understand, seeking data collected via practices they have used themselves (Zimmerman, 2007; Zimmerman, 2008) and with familiar data types (Murillo, 2014). Contextual details are found in field notebooks (Weber et al., 2012) and articles (Carlson & Stowel-Bracke, 2013), but additional metadata attached to datasets are the preferred method of conveying context (Bowker, 2000b). Formal metadata has limitations, though, as they cannot always contain enough detail or inspire the confidence needed for reuse. Researchers may instead base decisions on the word-of-mouth reputation of the dataset (Weber et al., 2012) or rely on more experienced researchers to develop understanding or alternative evaluation strategies (Zimmerman, 2008)

Data must have sufficient quality, often defined by community standards, to be reused (Zimmerman, 2007). Water researchers and earth science modelers consider comprehensiveness and continuity over time and space (Dow et al., 2015; Parsons, 2011) as well as uncertainties and error estimates (Larsen, Hamilton, Lucido, Garner, & Young, 2016; Parsons, 2011) when determining data quality. Volcanologists use image resolution as a quality indicator (Weber et al., 2012).

Ecologists trust data from well-known sources, such as databases and literature (Zimmerman, 2007), and make decisions based on authors' reputations and affiliations (Murillo, 2014; Weber et al., 2012). Initial evaluations are based on the reputation of the source where the data were discovered, even if researchers eventually obtain them through other means (Zimmerman, 2007). Standardized collection practices are not enough to establish trust, as practices themselves say nothing about the data collector's skill (Zimmerman, 2008). The sponsor of research (McNie, 2007) and membership in the same community of practice (Van House et al., 1998) facilitate trust among environmental planners and policy makers.

Both ecologists and modelers reuse data that are easy to access (Zimmerman, 2007; Parsons, 2011). Modelers, however, face an extra step in the evaluation process, needing first to decode numerically encoded datasets before deciding if they are appropriate (Pallickara et al., 2010).

**Biomedicine**

Visual, medical, and textual criteria are used to evaluate biomedical images. Health care workers rank visual relevance, background information, and image quality as being most important, although they also mention image modality and understandability (Clough, Sedghi, & Sanderson, 2008). Radiologists rely on a mixture of image properties, image quality, supporting documentation, and information about the source to determine suitability (Markonis et al., 2012).

Evaluation criteria vary depending on users' professional specialties and particular situations (Clough et al., 2008). Users rely on visual attributes when evaluating general medical images but incorporate textual information and credibility criteria for specific images used for background purposes (Sedghi, Sanderson, & Clough, 2011).

Definitions of quality also vary by user. A neurosurgeon, for example, uses noise levels, resolution, and anatomical coverage, while a radiologist focuses mostly on motion artefacts to determine image quality (Heckel, Arlt, Geisler, Zidowitz, & Neumuth, 2016). Resolution and acquisition details (e.g. slice thickness in tomographic images) are other proxies for quality (Müller et al., 2006).

Healthcare professionals determine relevance through a combination of textual background information, visual inspection, and mental comparison to imagined ideals (Sedghi, Sanderson, & Clough, 2012). Personal experience trumps other criteria, however, when determining image relevance (Markonis et al., 2012; Müller et al., 2006).

Clinicians build trust in images through supporting documentation, such as attached exams or biopsies. Systems allowing researchers to comment on images online can also build trust normally created through informal "hallway" communications (Jirotka et al., 2005; Markonis et al., 2012).

**Field Archaeology**

Archaeologists require contextual information about collection methods, instrumentation, observational conditions, and artefact provenance (Faniel, Barrera-Gomez, et al., 2013). Other fundamental metadata include information about original research goals and baseline geographic, stratigraphic, and chronological data (Atici et al., 2013). Current metadata schemas are not rich enough to provide this level of contextual description. Archaeologists either make do with the available information or seek other ways to further develop context (Faniel, Kansa, et al., 2013).

Consistent data recording practices (e.g. an absence of misspellings or translational errors) (Atici et al., 2013), and detailed language in supporting documentation (Faniel, Kansa, et al., 2013) help to establish credibility and trustworthiness. Author reputation and affiliation and repository features, such as metadata type and level of transparency, help to establish trust (Faniel, Kansa, et al., 2013)

**Social Sciences**

DBK users spend more time evaluating data results compared to literature results, consulting additional documentation when needed. Researchers appear to think this is normal, perhaps because choosing the correct dataset is more important than selecting the correct article (Kern & Mathiak, 2015). Title and publication year are not as important as study time frame and keywords in evaluations. Users would like access to literature citing a dataset to determine if a research question has already been answered (Kern & Mathiak, 2015); prior reuse of data is also an important way of developing trust (Faniel & Yakel, 2017).

Data seekers rank accessibility as the most important factor determining satisfaction with data reuse in the ICPSR repository. Data completeness (ranked 2$^{nd}$), credibility (4$^{th}$) and ease of use (5$^{th}$) are also contributing factors; in this study, journal/author reputation do not appear to impact satisfaction (Faniel et al., 2016). Other work suggests that the repository reputation is an important signal of data quality and credibility (Curty, 2016) and is used to develop trust in data (Faniel & Yakel, 2017). Data re-users tend to either make do with available data or adapt their research projects to use data that they can find. The more researchers have to “reshape” their projects, the less satisfied they are (Faniel et al., 2016).

Users need contextual information about collection methods, instrumentation, other analyses, and how variables are defined and measured (Curty, 2016; Faniel, Kansa, et al., 2013; Kern & Mathiak, 2015; Yoon, 2014a). When necessary, researchers turn to other sources to develop the necessary context (Fielding & Fielding, 2008), consulting colleagues, codebooks (Faniel & Yakel, 2017) or bibliographies (Faniel, Barrera-Gomez, et al., 2013). Ideally, specialized metadata schemas would provide enhanced context (Kern & Mathiak, 2015). Debate remains, however, if documentation can build the context needed to reuse qualitative social science data (Broom et al., 2009; Parry & Mauthner, 2005).

Novice researchers especially need supporting contextual information. They want details about coding procedures, collection methods and dataset merging and matching (Faniel, Kriesberg, & Yakel, 2012). More experienced researchers can make do more easily with limited documentation (Yoon, 2016).

## Discussion

Having presented the documented practices of observational data users, we use the frameworks to synthesize our key findings and to identify commonalities and themes spanning the reviewed disciplinary communities.

**Users and Needs**

Researchers across and within the reviewed disciplines need a diversity of observational data, requiring data of different types from different sources and disciplines, collected at different scales using different instruments. Users have very specific requirements, needing data from particular locations (geographic, anatomical, or astronomical), at particular resolutions or collected using particular mechanical or survey instruments.

Integrating diverse data is necessary but challenging. Astronomers struggle to bring together data from different telescopes, neuroscientists try to combine neuroimages with clinical data, and archaeologists need to integrate data collected in different contexts with different methodologies. Some of these challenges may be augmented by changes in research practices, such as automated data collection in EES (Borgman et al., 2007), or by shifts in community data practices, such as increased data sharing, as in archaeology (Arbuckle et al., 2014) or neuroscience (Choudhurry et al., 2014).

Background and foreground uses are reported across disciplines, although background uses are better documented. These include making comparisons, benchmarking, preparing research projects, calibrating instruments, and as model inputs. Reported foreground uses are vaguer, often limited to reports of "asking new questions of data." This does not mean that foreground uses do not occur; examples of new research fueled by data reuse could likely be found in all of the reviewed disciplines (e.g. Atici et al., 2017). This could indicate a mismatch between what studies of data practices report and actual practices, or it could be a sign of changing practices. Even with a broad analysis, we see that data use varies within disciplines. One group of biodiversity researchers uses secondary data only to support projects, for example, while another study only examines cases of foreground use. Other possible data uses, i.e. in teaching, clinical practice, or environmental planning, are hinted at, although rarely explored in detail.

A generic view of the user is also common. Similar to our approach, disciplines are often broadly represented; the social sciences in particular tend to be treated as a homogenous group. Few studies document the needs and behaviors of specific user groups, such as early career researchers (Kriesberg et. al, 2013; Faniel et al. 2012), policy makers (Janse, 2006; McNie, 2007; Cash et. al, 2003) or students (Carlson & Stowel-Bracke, 2013). Understanding the data practices of ECRs sheds light on processes of acculturation (Kriesberg et al, 2013) and is important, as large-scale data reuse depends on adoption by ECRs (Faniel et al, 2012). Understanding the practices of specific user groups is also critical in designing user-oriented data discovery systems.

**User Actions**

Across communities, users find data in repositories, journals, on websites, and through personal networks. This variety could be due to differing infrastructures available within disciplines; however, even in fields with established data repositories, i.e. astronomy and quantitative social science, researchers seek data outside of these systems (Sands et al., 2012; Faniel & Yakel, 2017).

Personal exchanges are valuable sources of external data. While locating large, well-known datasets is straightforward, tracking down smaller, specialized datasets is challenging and often requires personal communication (Sands et al., 2012). Existing repository search functionalities may not meet the specific needs of researchers, or users may not develop appropriate search strategies in these resources (Sedghi et al., 2011). Users may also simply not be aware of the existence of data or databases; this may be especially true for researchers seeking data outside of their primary disciplines.

The distributed nature of observational data compounds these problems. A variety of data repositories exist within these disciplines (e.g. Dow et. al, 2015); within each new resource, users must start from scratch - first discovering the resource, then investing significant time and energy becoming familiar with it and the available data. A lack of time and accessible data also complicates the search process.

**Evaluation**

Researchers across disciplines need as much contextual information as possible, requiring documentation about instruments, methodologies, research questions, and observational conditions. This information is combined with the reputation of the repository and often that of the data author to establish trust, data quality, and relevance. Although much of the reviewed literature uses frames such as trust and quality to discuss evaluation, the characteristics used to develop these frames varies. This variation may result from disciplinary or individual differences or from how the papers' authors define these frames. One commonality that we can identify is the association of more social criteria - such as the reputation of authors and data sources - in developing trust.

Enriched metadata are often the desired, although imperfect, method of conveying contextual information. Perhaps because of limitations in metadata, researchers build the needed information by combining a variety of sources, from codebooks and academic literature to unpublished reports and museum records (Faniel & Yakel, 2017). Researchers across communities also use social connections and personal exchanges to evaluate data. The discussion about how researchers evaluate data is still developing, although the process seems to differ from how researchers evaluate academic literature.

The following themes bridging both frameworks emerge from this synthesis:

- A tension between breadth and specificity
- The social aspects of data retrieval
- Absent practices and communities

When developing the frameworks for this paper, we presented the tension involved in applying broad perspectives to understand individual practices. This tension between breadth and specificity is also present in the reviewed data retrieval practices. Even within disciplines, researchers need a diversity of observational data and employ a wide variety of search and evaluation strategies. At the same time, users seek data with very precise characteristics. They appear to balance breadth and specificity as they work to integrate datasets from diverse sources to meet specific needs or to piece together a variety of evaluation criteria to make decisions about reuse.

Social connections and personal exchanges permeate observational data retrieval. Users rely on personal connections and their own networks to locate, obtain, and evaluate data, even in disciplines with extensive infrastructures. This suggests that it is not enough to understand data retrieval as a series of interactions between users and search systems; rather, data retrieval is in fact a complex socio-technical process.

The absence of many communities and practices in the literature is also apparent. A relatively small number of disciplines are represented in our literature corpus. Among the broad disciplinary categories that we employ, certain subdisciplines are well represented; others are briefly mentioned, and others are treated homogeneously. Building a robust picture of observational data retrieval requires a deeper understanding of practices in other disciplines and of understudied user groups such as non-scientists or early career researchers. Deeper studies of how data retrieval practices change when seeking data for foreground purposes, or when seeking data from different disciplines, are also absent. Although Faniel & Yakel have recently identified five “trust markers” important in data reuse in archeology, social sciences and zoology (2017), common frameworks for discussing evaluation criteria across the observational data community are lacking.

## Conclusion: Towards a Model for Data Retrieval

Through our analysis we have achieved the following:

- Shown that a framework based on interactive information retrieval is applicable to understanding the data retrieval literature
- Tested the boundaries of defining data communities, using broad classifications to identify commonalities in practices
- Revealed absent practices and highlighted areas where more research is necessary
- Suggested that a framework based on IR alone is insufficient for completely understanding the complexity of data retrieval practices.

The literature also points to ways that information retrieval and data retrieval differ. Data needs are specific, requiring high precision in IR systems (Stempfhuber & Zapilko, 2009). Textual queries and ranking algorithms do not work well for retrieving numeric or encoded data (Pallickara et al., 2010). Users employ different search strategies when seeking data than literature (Kern & Mathiak, 2015) and take different roles when interacting with data repositories (e.g. as consumers and creators), which can impact system design (Borgman et al., 2015). Researchers also spend more time evaluating datasets (Kern & Mathiak, 2015), perhaps because lists of data cannot be efficiently evaluated in the same way as document lists (Kunze & Auer, 2013).

These differences, in conjunction with the themes identified in the discussion, suggest that current information retrieval models may not completely describe data retrieval practices. Identifying commonalities in observational data retrieval practices is a first step in exploring possible characteristics of a new model for data information retrieval. Further studies of different data communities, such as users of experimental and computational data, big and long-tail data seekers, and members of underrepresented user groups are needed. A model describing data retrieval would provide insight into the needs and practices of users that could be applied to both systems design and policy developments for facilitating data discovery.

## Footnotes

1. IR systems are also an important part of these models. The first version of the article preprint (arXiv:1707.06937) includes an additional review of data retrieval systems.

2. For a detailed methodology and machine-readable bibliography, including references regarding data retrieval systems, see: https://doi.org/10.17026/dans-zgu-qfpj

## Acknowledgements

KG developed the frameworks, collected the data, and wrote the manuscript. PG, HC, SW, and AS contributed to theory development and editing. This work was funded by the Netherlands Organization for Scientific Research, Grant 652.001.002.

## References

Aloia, N., Binding, C., Cuy, S., Doerr, M., Felicetti, A., Fihn, J., … Richards, J. (2017). Enabling European archaeological research: The ARIADNE e-Infrastructure. *Internet Archaeology*, *43*, 1–21.

Ames, D. P., Horsburgh, J. S., Cao, Y., Kadlec, J., Whiteaker, T., & Valentine, D. (2012). HydroDesktop: Web services-based software for hydrologic data discovery, download, visualization, and analysis. *Environmental Modelling and Software*, *37*, 146–156. https://doi.org/10.1016/j.envsoft.2012.03.013

Arbuckle, B. S., Kansa, S. W., Kansa, E., Orton, D., Çakırlar, C., Gourichon, L., ... & Buitenhuis, H. (2014). Data sharing reveals complexity in the westward spread of domestic animals across Neolithic Turkey. *PloS one*, *9*(6), e99845.

Atici, L., Kansa, S. W., Lev-Tov, J., & Kansa, E. C. (2013). Other people's data: A demonstration of the imperative of publishing primary data. *Journal of Archaeological Method and Theory*, *20*(4), 663–681. https://doi.org/10.1007/s10816-012-9132-9

Atici L, Pilaar Birch S.E., Erdoğu B. (2017). Spread of domestic animals across Neolithic western Anatolia: New zooarchaeological evidence from Uğurlu Höyük, the island of Gökçeada, Turkey. *PLoS ONE, 12*(10): e0186519. https://doi.org/10.1371/journal.pone.0186519

Bahls, D., & Tochtermann, K. (2013). Semantic retrieval interface for statistical research data. In *Proceedings of the 3rd International Workshop on Semantic Digital Archives* (pp. 93–103).

Baker, K. S., Duerr, R. E., & Parsons, M. A. (2015). Scientific knowledge mobilization: co-evolution of data products and designated communities. *International Journal of Digital Curation*, *10*(2), 110–135. https://doi.org/10.2218/ijdc.v10i2.346

Bates, M. (1990). Where should the person stop and the information search start? *Information, Processing and Management*, *26*(5), 575–591.

Beaulieu, A. (2004). From brainbank to database: The informational turn in the study of the brain. *Studies in History and Philosophy of Biological and Biomedical Sciences*, *35*(2), 367–390. https://doi.org/10.1016/j.shpsc.2004.03.011

Belkin, N. J. (1993). Interaction with texts: information retrieval as information-seeking behavior. In G. Knorz, J. Krause, & C. Womser-Hacker (Eds.), *Information retrieval' 93: Von der Modellierung zur Anwendung* (pp. 55–66). Konstanz: Universitaetsverlag Konstanz.

Belkin, N. J. (1996). Intelligent information retrieval: whose intelligence? In *ISI '96: Proceedings of the Fifth International Symposium for Information Science* (pp. 25–31). Konstanz: Universtaetsverlag Konstanz.

Beran, B., Cox, S. J. D., Valentine, D., Zaslavsky, I., & McGee, J. (2009). Web services solutions for hydrologic data access and cross-domain interoperability. *International Journal on Advances in Intelligent Systems*, *2*(2&3).

Birnholtz, J. P., & Bietz, M. J. (2003). Data at work: Supporting sharing in science and engineering. In *Proceedings of the International ACM SIGGROUP Conference on Supporting Group Work*.

Borgman, C. L. (2015). *Big data, little data, no data: Scholarship in the networked world*. MIT press.

Borgman, C. L., Darch, P. T., Sands, A. E., & Golshan, M. S. (2016). The durability and fragility of knowledge infrastructures: Lessons learned from astronomy. In *Proceedings of the 79th ASIS&T Annual Meeting*. Copenhagen. Retrieved from http://arxiv.org/abs/1611.00055

Borgman, C. L., Golshan, M. S., Sands, A. E., Wallis, J. C., Cummings, R. L., Darch, P., & Randies,

B. M. (2016). Data management in the long tail: Science, software, and service. *International Journal of Digital Curation*, *11*(1), 128–149. https://doi.org/10.2218/ijdc.v11i1.428
Borgman, C. L., Scharnhorst, A., & Golshan, M. S. (2018). Digital Data Archives as Knowledge Infrastructures: Mediating Data Sharing and Reuse. *arXiv preprint arXiv:1802.02689*.
Borgman, C. L., Smart, L. J., Millwood, K. A., Finley, J. R., Champeny, L., Gilliland, A. J., & Leazer, G. H. (2005). Comparing faculty information seeking in teaching and research: Implications for the design of digital libraries. *Journal of the American Society for Information Science and Technology*, *56*(6), 636–657. https://doi.org/10.1002/asi.20154
Borgman, C. L., Van de Sompel, H., Scharnhorst, A., van den Berg, H., & Treloar, A. (2015). Who uses the digital data archive? An exploratory study of DANS. In *Proceedings of the Association for Information Science and Technology* (Vol. 52). https://doi.org/10.1002/pra2.2015.145052010096
Borgman, C. L., Wallis, J. C., & Enyedy, N. (2007). Little science confronts the data deluge: Habitat ecology, embedded sensor networks, and digital libraries. *International Journal on Digital Libraries*, *7*(1–2), 17–30. https://doi.org/10.1007/s00799-007-0022-9
Bowker, G. C. (2000a). Biodiversity datadiversity. *Social Studies of Science*, *30*(5), 643–83.
Bowker, G. C. (2000b). Work and information practices in the sciences of biodiversity. In *Proceedings of the 26th International Conference on Very Large Data Bases*. Cairo, Egypt.
Broom, A., Cheshire, L., & Emmison, M. (2009). Qualitative researchers' understandings of their practice and the implications for data archiving and sharing. *Sociology*, *43*(6), 1163–1180. https://doi.org/10.1177/0038038509345704
Brown, C. (2003). The changing face of scientific discourse: Analysis of genomic and proteomic database usage and acceptance. *Journal of the American Society for Information Science and Technology*, *54*(10), 926–938. https://doi.org/10.1002/asi.10289
Carlson, J., & Stowel-Bracke, M. (2013). Data management and sharing from the perspective of graduate students: An examination of the culture and practice at the water quality field station. *Libraries Faculty and Staff Scholarship and Research*. https://doi.org/10.1353/pla.2013.0034
Cash, D. W., Clark, W. C., Alcock, F., Dickson, N. M., Eckley, N., Guston, D. H., … Mitchelll, R. B. (2003). Knowledge systems for sustainable development. *Proceedings of the National Academy of Sciences of the United States of America*, *100*(14), 8086–8091. https://doi.org/10.1073/pnas.1231332100
Choudhury, S., Fishman, J. R., McGowan, M. L., & Juengst, E. T. (2014). Big data, open science and the brain: Lessons learned from genomics. *Frontiers in Human Neuroscience*, *8*, 1–10. https://doi.org/10.3389/fnhum.2014.00239
Clough, P. D., Sedghi, S., & Sanderson, M. (2008). A study on the relevance criteria for medical images. *Pattern Recognition Letters*, *29*(15), 2046–2057. https://doi.org/10.1186/1742-7622-5-2
Corti, L. (2007). Re-using archived qualitative data - Where, how, why? *Archival Science*, *7*(1), 37–54. https://doi.org/10.1007/s10502-006-9038-y
Cragin, M. H., Chao, T. C., & Palmer, C. L. (2011). Units of evidence for analyzing subdisciplinary difference in data practice studies. In *Proceedings of the 11th ACM/IEEE-CS Joint Conference on Digital Libraries*. https://doi.org/10.1145/1998076.1998175
Curty, R. G. (2016). Factors influencing research data reuse in the social sciences: An exploratory study. *International Journal of Digital Curation*, *11*(1), 96–117. https://doi.org/DOI: 10.2218/ijdc.v11i1.401
Davis, M. L. E. S., Tenopir, C., Allar, S., & Frame, M. T. (2014). Facilitating access to biodiversity information: A survey of users' needs and practices. *Environmental Management*, *53*(3), 690–701. https://doi.org/10.1007/s00267-014-0229-7
De-Arteaga, M., Eggel, I., Do, B., Rubin, D., Kahn, C. E., & Müller, H. (2015). Comparing image search behaviour in the ARRS GoldMiner search engine and a clinical PACS/RIS. *Journal of Biomedical Informatics*, *56*, 57–64. https://doi.org/10.1016/j.jbi.2015.04.013
Devarakonda, R., Palanisamy, G., Green, J. M., & Wilson, B. E. (2011). Data sharing and retrieval using OAI-PMH. *Earth Science Informatics*, *4*(1), 1–5. https://doi.org/10.1007/s12145-010-0073-0
Dow, A. K., Dow, E. M., Fitzsimmons, T. D., & Materise, M. M. (2015). Harnessing the environmental data flood: A comparative analysis of hydrologic, oceanographic, and meteorological informatics platforms. *Bulletin of the American Meteorological Society*, *96*(5), 725–736. https://doi.org/10.1175/BAMS-D-13-00178.1
Dulisch, N., Kempf, A. O., & Schaer, P. (2015). Query expansion for survey question retrieval in the social sciences. In S. Kapidakis, C. Mazurek, & M. Werla (Eds.), *Research and Advanced Technology for Digital Libraries. Lecture Notes in Computer Science* (Vol. 9316). Springer.

https://doi.org/10.1007/978-3-319-24592-8_3
Edwards, P. N., Mayernik, M. S., Batcheller, A. L., Bowker, G. C., & Borgman, C. L. (2011). Science friction: Data, metadata, and collaboration. *Social Studies of Science*, *41*(5), 667–690. https://doi.org/10.1177/0306312711413314
Erinjeri, J. P., Picus, D., Prior, F. W., Rubin, D. A., & Koppel, P. (2009). Development of a google-based search engine for data mining radiology reports. *Journal of Digital Imaging*, *22*(4), 348–356. https://doi.org/10.1007/s10278-008-9110-7
Faniel, I. M., Barrera-Gomez, J., Kriesberg, A., & Yakel, E. (2013). A comparative study of data reuse among quantitative social scientists and archaeologists. In *iConference 2013 Proceedings* (pp. 797–800). https://doi.org/10.9776/13391
Faniel, I. M., Kansa, E., Kansa, S. W., Barrera-Gomez, J., & Yakel, E. (2013). The challenges of digging data: A study of context in archaeological data reuse. In *Proceedings of the 13th ACM/IEEE-CS Joint Conference on Digital Libraries* (pp. 295–304). New York, NY: ACM. https://doi.org/http://dx.doi.org/10.1145/2467696.2467712
Faniel, I. M., Kriesberg, A., & Yakel, E. (2012). Data reuse and sensemaking among novice social scientists. *Proceedings of the ASIST Annual Meeting*, *49*(1). https://doi.org/10.1002/meet.14504901068
Faniel, I. M., Kriesberg, A., & Yakel, E. (2016). Social scientists' satisfaction with data reuse. *Journal of the Association for Information Science and Technology*, *67*(6), 1404–1416. https://doi.org/10.1002/asi.23480
Faniel, I. M., & Yakel, E. (2017). Practices do not make perfect: Disciplinary data sharing and reuse practices and their implications for repository data curation. *Curating Research Data, Volume One: Practical Strategies for Your Digital Repository*, 103-126.
Faniel, I. M., & Zimmerman, A. (2011). Beyond the data deluge: A research agenda for large-scale data sharing and reuse. *International Journal of Digital Curation*, *6*(1), 58–69. https://doi.org/10.2218/ijdc.v6i1.172
Fielding, N. G., & Fielding, J. L. (2008). Resistance and adaptation to criminal identity: Using secondary analysis to evaluate classic studies of crime and deviance. *Historical Social Research*, *33*(3), 75–93. https://doi.org/10.1177/S0038038500000419
Gray, J. (2009). Jim Gray on eScience: A transformed scientific method. In T. Hey, S. Tansley, & K. Tolle (Eds.), *The Fourth Paradigm: Data-Intensive Scientific Discovery* (pp. xvii–xxxi). Richmond, WA: Microsoft Research.
Gregory, K., Cousijn, H., Groth, P., Scharnhorst, A., & Wyatt, S. (2018). Understanding Data Retrieval Practices: A Social Informatics Perspective. *arXiv preprint arXiv:1801.04971*
Heckel, F., Arlt, F., Geisler, B., Zidowitz, S., & Neumuth, T. (2016). Evaluation of image quality of MRI data for brain tumor surgery. In *Proc. SPIE 9787, Medical Imaging 2016: Image Perception, Observer Performance, and Technology Assessment* (Vol. 9787). https://doi.org/10.1117/12.2214944
Hersh, W., Müller, H., Gorman, P., & Jensen, J. (2005). Task analysis for evaluating image retrieval systems in the ImageCLEF biomedical image retrieval task. In *Slice of Life Conference on Multimedia in Medical Education*. Portland, Oregon.
Hoeppe, G. (2014). Working data together: The accountability and reflexivity of digital astronomical practice. *Social Studies of Science*, *44*(2), 243–270. https://doi.org/10.1177/0306312713509705
Honor, L. B., Haselgrove, C., Frazier, J. A., & Kennedy, D. N. (2016). Data citation in neuroimaging: Proposed best practices for data identification and attribution. *Frontiers in Neuroinformatics*, *10*, 1–12. https://doi.org/10.3389/fninf.2016.00034
Ingwersen, P. (1992). *Information retrieval interaction*. London: Taylor Graham.
Ingwersen, P. (1996). Cognitive perspectives of information retrieval interaction: elements of a cognitive IR theory. *Journal of Documentation*, *52*(1), 3–50.
Janse, G. (2006). Information search behaviour of European forest policy decision-makers. *Forest Policy and Economics*, *8*(6), 579–592.
Jirotka, M., Procter, R., Hartswood, M., Slack, R., Simpson, A., Coopmans, C., … Voss, A. (2005). Collaboration and trust in healthcare innovation: The eDiaMoND case study. *Computer Supported Cooperative Work:*, *14*(4), 369–398. https://doi.org/10.1007/s10606-005-9001-0
Kalpathy-Cramer, J., Herrera, A. G. S. de, Demner-Fushman, D., Antani, S., Bedrick, S., & Müller, H. (2015). Evaluating performance of biomedical image retrieval systems – an overview of the medical image retrieval task at ImageCLEF 2004–2013. *Computerized Medical Imaging and Graphics*, *39*, 55–61. https://doi.org/10.1016/j.compmedimag.2014.03.004
Karcher, S., Kirilova, D., & Weber, N. (2016). Beyond the matrix: Repository services for qualitative data. *IFLA Journal*, *42*(4). https://doi.org/10.1177/0340035216672870

Kern, D., & Mathiak, B. (2015). Are there any differences in data set retrieval compared to well-known literature retrieval? In S. Kapidakis, C. Mazurek, & M. Werla (Eds.), *Research and Advanced Technology for Digital Libraries. Lecture Notes in Computer Science* (Vol. 9316). Springer, Cham. https://doi.org/10.1007/978-3-319-24592-8_15

Key Perspectives. (2010). *Data dimensions: Disciplinary differences in research data sharing, reuse and long term viability. SCARP Synthesis Study*. Digital Curation Centre. Retrieved from http://www.dcc.ac.uk/scarp

Kim, S., & Gilbertson, J. (2007). Information requirements of cancer center researchers focusing on human biological samples and associated data. *Information Processing and Management*, *43*(5), 1383–1401. https://doi.org/10.1016/j.ipm.2006.10.012

Kriesberg, A., Frank, R. D., Faniel, I. M., & Yakel, E. (2013). The role of data reuse in the apprenticeship process. *Proceedings of the ASIST Annual Meeting*, *50*(1). https://doi.org/10.1002/meet.14505001051

Kuhlthau, C. C. (1991). Inside the search process: Information seeking from the user 's perspective. *Journal of the American Society for Information Science*, *42*(5), 361–371. https://doi.org/10.1002/(SICI)1097-4571(199106)42:5<361::AID-ASI6>3.0.CO;2-#

Kunze, S. R., & Auer, S. (2013). Dataset retrieval. In *2013 IEEE 7th International Conference on Semantic Computing, ICSC 2013* (pp. 1–8). https://doi.org/10.1109/ICSC.2013.12

Larsen, S., Hamilton, S., Lucido, J., Garner, B., & Young, D. (2016). Supporting diverse data providers in the open water data initiative: Communicating water data quality and fitness of use. *Journal of the American Water Resources Association*, *52*(4), 859–872. https://doi.org/10.1111/1752-1688.12406

Leonelli, S. (2016). *Data-centric biology: a philosophical study*. University of Chicago Press.

Maier, D., Megler, V. M., Baptista, A. M., Jaramillo, A., Seaton, C., & Turner, P. J. (2012). Navigating oceans of data. In A. Ailamaki & S. Bowers (Eds.), *Scientific and Statistical Database Management. SSDBM 2012. Lecture Notes in Computer Science* (Vol. 7338, pp. 1–19). Springer. https://doi.org/10.1007/978-3-642-31235-9_1

Maier, D., Megler, V. M., & Tufte, K. (2014). Challenges for dataset search. In S. S. Bhowmick, C. E. Dyreson, C. S. Jensen, M. L. Lee, A. Muliantara, & B. Thalheim (Eds.), *Database Systems for Advanced Applications. DASFAA 2014. Lecture Notes in Computer Science* (Vol. 8421). Springer. https://doi.org/10.1007/978-3-319-05810-8_1

Markonis, D., Holzer, M., Dungs, S., Vargas, A., Langs, G., Kriewel, S., & Müller, H. (2012). A survey on visual information search behavior and requirements of radiologists. *Methods of Information in Medicine*, *51*(6), 539–548. https://doi.org/10.3414/ME11-02-0025

McNie, E. C. (2007). Reconciling the supply of scientific information with user demands: an analysis of the problem and review of the literature. *Environmental Science and Policy*, *10*(1), 17–38. https://doi.org/10.1016/j.envsci.2006.10.004

Megler, V. M., & Maier, D. (2012). When big data leads to lost data. In *Proceedings of the 5th Ph.D. workshop on Information and knowledge* (pp. 1–8). Maui, Hawaii: ACM. https://doi.org/10.1145/2389686.2389688

Meyer, E. T., Bulger, M., Kyriakidou-Zacharoudiou, A., Power, L., Williams, P., Venters, W., … Wyatt, S. (2011). *Collaborative yet independent: information practices in the physical sciences*. London: Research Information Network.

Michener, W. K. (2015). Ecological data sharing. *Ecological Informatics*, *29*, 33–44. https://doi.org/10.1016/j.ecoinf.2015.06.010

Miller, J. R., Groom, M., Hess, G. R., Steelman, T., Stokes, D. L., Thompson, J., … Marquardt, R. (2009). Biodiversity conservation in local planning. *Conservation Biology*, *23*(1), 53–63.

Müller, H., Despont-Gros, C., Hersh, W., Jensen, J., Lovisa, C., & Antoine Geissbuhler. (2006). Health care professionals' image use and search behaviour. In *Proceedings of the Medical Informatics Europe Conference (MIE 2006)* (pp. 24–32). Maastricht, Netherlands: IOS Press, Studies in Health Technology and Informatics.

Murillo, A. P. (2014). Examining data sharing and data reuse in the DataONE environment. *Proceedings of the ASIST Annual Meeting*, *51*(1). https://doi.org/10.1002/meet.2014.14505101155

National Science Board. (2005). *Long-lived digital data collections: Enabling research and education in the 21st century*. National Science Foundation. Retrieved from https://www.nsf.gov/pubs/2005/nsb0540/nsb0540.pdf

National Science Foundation. (2007). *Cyberinfrastructure vision for 21st century discovery*. Retrieved from https://www.nsf.gov/pubs/2007/nsf0728/nsf0728.pdf

Niccolucci, F., & Richards, J. D. (2013). ARIADNE: Advanced Research Infrastructures for Archaeological Dataset Networking in Europe. *International Journal of Humanities and Arts Computing*, *7*(1–2), 70–88. https://doi.org/10.3366/ijhac.2013.0082

Pallickara, S. L. S., Pallickara, S. S., & Zupanski, M. (2012). Towards efficient data search and subsetting of large-scale atmospheric datasets. *Future Generation Computer Systems*, *28*(1), 112–118. https://doi.org/10.1016/j.future.2011.05.010

Pallickara, S. L. S., Pallickara, S. S., Zupanski, M., & Sullivan, S. (2010). Efficient metadata generation to enable interactive data discovery over large-scale scientific data collections. In *Proceedings of the 2010 IEEE Second International Conference on Cloud Computing Technology and Science* (pp. 573–580). IEEE Computer Society. https://doi.org/10.1109/CloudCom.2010.99

Palmer, C. L., Cragin, M. H., & Hogan, T. P. (2004). Information at the intersections of discovery: Case studies in neuroscience. In *Proceedings of the ASIST Annual Meeting* (Vol. 41, pp. 448–455). https://doi.org/10.1002/meet.1450410152

Parry, O., & Mauthner, N. (2005). Back to basics: Who re-uses qualitative data and why? *Sociology*, *39*(2), 337–342. https://doi.org/10.1177/0038038505050543

Parsons, M. A. (2011). Making data useful for modelers to understand complex earth systems. *Earth Science Informatics*, *4*, 197–223. https://doi.org/10.1007/s12145-011-0089-0

Pasquetto, I. V., Randles, B. M., & Borgman, C. L. (2017). On the reuse of scientific data. *Data Science Journal*, *16*(8), 1–9. https://doi.org/10.5334/dsj-2017-008

Pepe, A., Goodman, A., Muench, A., Crosas, M., & Erdmann, C. (2014). How do astronomers share data? Reliability and persistence of datasets linked in AAS publications and a qualitative study of data practices among US astronomers. *PLoS ONE*, *9*(8). https://doi.org/10.1371/journal.pone.0104798

Pullin, A., Knight, T., Stone, D., & Charman, K. (2004). Do conservation managers use scientific evidence to support their decision making? *Biological Conservation*, *119*(2), 245–252.

Rieh, S. Y., & Xie, H. (2006). Analysis of multiple query reformulations on the web: The interactive information retrieval context. *Information Processing and Management*, *42*(3), 751–768. https://doi.org/10.1016/j.ipm.2005.05.005

Sanderson, M., & Croft, W. B. (2012). The history of information retrieval research. *Proceedings of the IEEE*, *100*, 1444–1451. https://doi.org/10.1109/JPROC.2012.2189916

Sands, A., Borgman, C. L., Wynholds, L., & Traweek, S. (2012). Follow the data: How astronomers use and reuse data. In *Proceedings of the ASIST Annual Meeting* (Vol. 49, pp. 1–3). Baltimore, MD. https://doi.org/10.1002/meet.14504901341

Saracevic, T. (1996). Modeling interaction in information retrieval (IR): a review and proposal. In *Proceedings of the 59th Annual Meeting of the American Society for Information Science* (pp. 3–9).

Saracevic, T. (1997). The stratified model of information retrieval interaction: extension and applications. In *Proceedings of the 60th Annual Meeting of the American Society for Information Science* (pp. 313–327).

Schmidt, B., Gemeinholzer, B., & Treloar, A. (2016). Open data in global environmental research: The Belmont Forum's open data survey. *PLoS ONE*, *11*(1). https://doi.org/10.1371/journal.pone.0146695

Sedghi, S., Sanderson, M., & Clough, P. (2011). Medical image resources used by health care professionals. *Aslib Proceedings: New Information Perspectives*, *63*(6), 570–585. https://doi.org/10.1108/00012531111187225

Sedghi, S., Sanderson, M., & Clough, P. (2012). How do health care professionals select medical images they need? *Aslib Proceedings*, *64*(4), 437–456. https://doi.org/10.1108/00012531211244815

Shen, Y. (2007). Information seeking in academic research: A study of the sociology faculty at the University of Wisconsin-Madison. *Information technology and libraries*, *26*(1), 4.

Stempfhuber, M., & Zapilko, B. (2009). Integrated retrieval of research data and publications in digital libraries. In *Rethinking Electronic Publishing: Innovation in Communication Paradigms and Technologies - Proceedings of the 13th International Conference on Electronic Publishing* (pp. 613–620). Milano, Italy.

Swan, A., & Brown, S. (2008). *To share or not to share: Publication and quality assurance of research data outputs*. Retrieved from http://www.rin.ac.uk/system/files/attachments/To-share-data-outputs-report.pdf

Tenopir, C., Allard, S., Douglass, K., Aydinoglu, A. U., Wu, L., Read, E., … Frame, M. (2011). Data sharing by scientists: Practices and perceptions. *PLoS ONE*, *6*(6), e21101. https://doi.org/10.1371/journal.pone.0021101

Tenopir, C., Dalton, E. D. E. D., Allard, S., Frame, M., Pjesivac, I., Birch, B., … Dillman, D. (2015). Changes in data sharing and data reuse practices and perceptions among scientists worldwide. *PLoS ONE*, *10*(8), e0134826. https://doi.org/10.1371/journal.pone.0134826

Van Horn, J. D., & Gazzaniga, M. S. (2013). Why share data? Lessons learned from the fMRIDC. *NeuroImage*, *82*, 677–682. https://doi.org/10.1016/j.neuroimage.2012.11.010

Van House, N. A., Butler, M. H., & Schiff, L. R. (1998). Cooperative knowledge work and practices of trust: Sharing environmental planning data sets. In *Proceedings of the 1998 ACM Conference on Computer Supported Cooperative Work*. ACM.

Wallis, J. C., Borgman, C. L., Mayernik, M. S., Pepe, A., Ramanathan, N.., & Hansen, M. (2007). Know thy sensor: Trust, data quality, and data integrity in scientific digital libraries. In L. Kovács, N. Fuhr, & C. Meghini (Eds.), *Research and Advanced Technology for Digital Libraries. ECDL 2007. Lecture Notes in Computer Science* (Vol. 4675, pp. 380–391). Springer, Berlin, Heidelberg. https://doi.org/10.1007/978-3-540-74851-9_32

Wallis, J. C., Rolando, E., & Borgman, C. L. (2013). If we share data, will anyone use them? Data sharing and reuse in the long tail of science and technology. *PLoS ONE*, *8*(7). https://doi.org/10.1371/journal.pone.0067332

Weber, N. M., Baker, K. S., Thomer, A. K., Chao, T. C., & Palmer, C. L. (2012). Value and context in data use: Domain analysis revisited. In *Proceedings of the ASIST Annual Meeting* (Vol. 49). https://doi.org/10.1002/meet.14504901168

Weller, T., & Monroe-Gulick, A. (2014). Understanding methodological and disciplinary differences in the data practices of academic researchers. *Library Hi Tech*, *32*(3), 467. https://doi.org/10.1108/LHT-02-2014-0021

Williams, R., Pryor, G., Bruce, A., Macdonald, S., Marsden, W., Calvert, J., … Neilson, C. (2009). *Patterns of information use and exchange: Case studies of researchers in the life sciences*. Research Information Network.

Wolfram, D. (2015). The symbiotic relationship between information retrieval and informetrics. *Scientometrics*, *102*(3), 2201–2214. https://doi.org/10.1007/s11192-014-1479-0

Wynholds, L. A., Wallis, J. C., Borgman, C. L., Sands, A., & Traweek, S. (2012). Data, data use, and scientific inquiry: Two case studies of data practices. In *Proceedings of the 12th ACM/IEEE-CS Joint Conference on Digital Libraries* (pp. 19–22). https://doi.org/10.1145/2232817.2232822

Wynholds, L., Fearon, D. S., Borgman, C. L., & Traweek, S. (2011). When use cases are not useful: Data practices, astronomy, and digital libraries. In *Proceedings of the 11th Annual International ACM/IEEE Joint Conference on Digital Libraries* (pp. 383–386). https://doi.org/10.1145/1998076.1998146

Xie, I. (2008). *Interactive Information Retrieval in Digital Environments*. Hershey, PA: IGI Publishing.

Yoon, A. (2014a). End users' trust in data repositories: Definition and influences on trust development. *Archival Science*, *14*(1), 17–34. https://doi.org/10.1007/s10502-013-9207-8

Yoon, A. (2014b). "Making a square fit into a circle": Researchers' experiences reusing qualitative data. In *Proceedings of the ASIST Annual Meeting* (Vol. 51). https://doi.org/10.1002/meet.2014.14505101140

Yoon, A. (2016). Red flags in data: Learning from failed data reuse experiences; Red flags in data: Learning from failed data reuse experiences. In *Proceedings of the Association for Information Science and Technology* (pp. 1–6). https://doi.org/10.1002/pra2.2016.14505301126

Zimmerman, A. (2007). Not by metadata alone: The use of diverse forms of knowledge to locate data for reuse. *International Journal on Digital Libraries*, *7*(1–2). https://doi.org/10.1007/s00799-007-0015-8

Zimmerman, A. S. (2003). *Data sharing and secondary use of scientific data: Experiences of ecologists* (Unpublished). Ann Arbor: The University of Michigan.

Zimmerman, A. S. (2008). New knowledge from old data: The role of standards in the sharing and reuse of ecological data. *Science Technology and Human Values*, *33*(5), 631–652. https://doi.org/10.1177/0162243907306704

Zinzi, A., Capria, M. T., Palomba, E., Giommi, P., & Antonelli, L. A. (2016). MATISSE: A novel tool to access, visualize and analyze data from planetary exploration missions. *Astronomy and Computing*, *15*, 16-28.